# Java-Based Client-Server Application: Design and Implementation

Omkar Patil and Aarya Shirbhate

COMP1549: Advanced Programming
University of Greenwich
Old Royal Naval College
United Kingdom

*Abstract*— **This report presents the development of a networked distributed system utilizing socket programming in a distributed computing environment. The system, named Group Communication System (GCS), was implemented using Java to exemplify the principles of socket programming and communication protocols. GCS facilitates group-based client-server communication through command-line interface (CLI) interactions, allowing members to join a group, communicate, and manage group state seamlessly. The project emphasizes key programming principles such as fault tolerance, design patterns, and version control system (VCS) utilization. The report provides a comprehensive overview of the system architecture, implementation details, and practical considerations, offering insights into the technical background and operational aspects of distributed systems.**

## I. Introduction

In today's digital landscape, the essence of networked distributed systems resonates profoundly. These systems serve as the backbone of modern communication infrastructures, fostering seamless interaction and data interchange across diverse platforms and devices. They are instrumental in meeting the burgeoning demand for interconnectedness and real-time communication in our increasingly digital world. The impetus behind delving into networked distributed systems lies in unlocking their potential to develop robust, efficient, and scalable solutions that cater to the dynamic needs of contemporary society.

This report is a culmination of our exploration into networked distributed systems through the lens of a network-based chat application. Our coursework focused on developing a robust system that facilitates communication among users through a central server and clients. The application boasts features like broadcasting, private messaging, and allows clients to join through the server's IP address and port. Our approach involved leveraging Java sockets to establish seamless communication channels between clients and the server, ensuring efficient data transfer and interaction. Additionally, we implemented a dedicated ClientHandler class to streamline client-side operations and enhance system performance.

The structure of this report is designed to provide a comprehensive understanding of our system's design, implementation, and findings. We commence by delving into the intricacies of system design and the methodologies employed during implementation. Subsequently, we present a detailed analysis and discussion of the results obtained through rigorous testing and evaluation. Finally, we draw conclusions based on our findings and outline potential avenues for future research and development.

For access to the complete project repository, please visit: [COMP1549-JAVA-CW-GROUP_94]

## II. Design/implementation

For the implementation and design of the networked distributed system, we utilized Java and socket programming to establish communication channels between clients and the server. The environment for the implementation included Java Development Kit (JDK) for Java programming and standard Java libraries for socket communication.

**Implementation of Technical Requirements:**

*Client-Server Architecture*: The system follows a client-server architecture, where the server acts as a centralized entity managing communication among multiple clients. We implemented this architecture using Java's Socket and ServerSocket classes to establish connections and facilitate data exchange between clients and the server.

*ClientHandler Threads*: Each client connection is managed by a separate ClientHandler thread, responsible for handling client interactions and message processing. We implemented this functionality to enable concurrent handling of multiple clients and ensure optimal responsiveness of the system.

*TCP Communication Protocol*: The system employs TCP (Transmission Control Protocol) for reliable and ordered data transmission between clients and the server. TCP sockets are utilized to establish and maintain connections, ensuring the delivery of messages without loss or corruption.

*Fault Tolerance Mechanisms*: In the event of client failures or disconnections, the system incorporates fault tolerance mechanisms to handle such scenarios gracefully. When a client abruptly disconnects, the associated ClientHandler thread is terminated to prevent disruptions to the communication flow. Additionally, a designated coordinator client oversees group communications, ensuring continuity in the absence of the coordinator.

*Message Broadcasting:* The system supports both public and private message broadcasting. Public messages are broadcasted to all connected clients, facilitating group communication. Private messages, prefixed with "@" symbol, are directed to specific clients, ensuring targeted communication between sender and recipient.

*Request Member Details Feature:* Clients could request member details from the coordinator by using "@coordinatorID /memberdetails". Upon receiving a request, the server retrieves information about all connected clients, including their client IDs, IP addresses, and ports. This feature enhances the system's transparency and enables clients to obtain comprehensive information about other participants in the communication network.

*Quit Feature:* Allows clients to disconnect from the server using the "/quit" command. Upon receiving this command, the associated ClientHandler thread terminates, and the client is removed from the list of connected clients. This feature ensures that clients can exit the system without causing disruptions to other connected clients.

**Implementation Choices:**
*Design Patterns*:

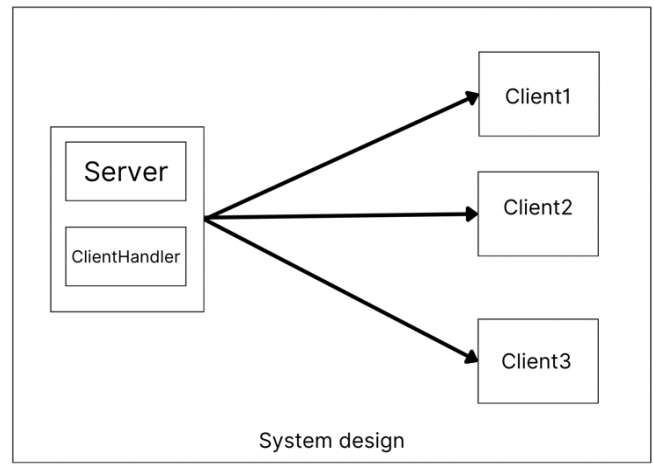
System design

*Singleton Pattern*: To ensure that only one instance of the server exists throughout the application's lifecycle, we employed the Singleton pattern. This design choice simplifies resource management and enhances system reliability by preventing the creation of multiple server instances.

*Factory Pattern:* The Factory pattern was utilized to dynamically generate ClientHandler objects, allowing for efficient management of client connections and promoting code modularity and reusability. This design choice facilitates the seamless integration of new features and extensions in the future.

*JUnit Tests:* Throughout the implementation, critical components of the system were subjected to rigorous JUnit testing to ensure their functionality and integration. Notable tests include **testServerIpandPort**, which verifies the server's IP address and port configuration, and **testgetCurrentTimestamp,** which validates the real-time timestamp generation capability. These tests, among others, played a pivotal role in identifying and rectifying potential issues early on, bolstering the reliability and robustness of the Group Communication System.

**Justification:**
The chosen design patterns and implementation choices were justified based on their effectiveness in achieving the desired system requirements. The Singleton pattern ensures the consistent behavior of the server instance, while the Factory pattern enhances modularity and extensibility. Additionally, the use of TCP communication protocol ensures reliable data transmission, and fault tolerance mechanisms improve the system's resilience in handling client failures or disconnections.

Overall, the implementation reflects a careful consideration of design principles, implementation choices, and testing strategies to deliver a robust, scalable, and fault-tolerant networked distributed system. The inclusion of features such as message broadcasting, member detail requests, and the quit feature enhances the system's functionality and usability, making it well-suited for diverse communication scenarios.

## III. Analysis and Critical Discussion

The analysis and debate of the Java-based client-server program demonstrate its strong features and efficiency in promoting user-to-user communication. All technical tasks are implemented effectively by the system, guaranteeing accurate message delivery to customers and seamless operation of private messaging features that facilitate focused communication. Additionally, the server carefully records every interaction together with the relevant time stamps, guaranteeing responsibility and openness in the chat room. This logging feature makes troubleshooting and behavior monitoring of the system easier in addition to offering a historical record of messages.

One of the notable strengths of the system lies in its fault tolerance mechanisms, which address both expected and unexpected client disconnections. When a typical client leaves with the "/quit" command, the system lets them go with grace and doesn't break up their conversation. The system excels at handling sudden client disconnections, though, continually tracking client connections and dynamically modifying its behavior to keep communication flowing. Through early detection and mitigation of client disturbances, the system improves user experience and builds trust and stability.

Modularity is a key principle guiding the system's design, achieved through the cohesive integration of various components. The TCP socket acts as the foundation for client-server communication, offering a dependable and effective means of exchanging data. Code reuse and maintainability are encouraged while clear and efficient communication is ensured via the client handler component, which oversees client-side actions. Because each component can be tested and optimized separately, the modular architecture not only makes the system more scalable and extensible but also makes debugging and troubleshooting easier.

The system's robustness and reliability are further validated through comprehensive JUnit testing. Tests such as "testServerIpandPort" verify the accuracy of server IP and port configurations, critical for establishing client-server connections. Similarly, the "testgetcurrenttimestamp" ensures that messages are timestamped correctly, enabling accurate tracking of communication events. Despite these strengths, the testing process also uncovers certain limitations, such as occasional false positives indicating server IP and port conflicts and delays in error message delivery due to network latency. These findings underscore the importance of ongoing testing and refinement to ensure optimal system performance and user satisfaction.

In conclusion, the analysis highlights the system's effectiveness in facilitating communication within a distributed environment while identifying opportunities for further enhancement and optimization. The system exhibits the capacity to accommodate a wide range of user communication requirements while upholding robustness and dependability under multiple operating conditions. This is achieved by utilizing fault tolerance mechanisms, modular design principles, and stringent testing processes.

## IV. Conclusions

In conclusion, the development and analysis of the networked distributed system presented in this study underscore the importance of robust communication infrastructure and fault-tolerant design in facilitating seamless interaction among clients. By adopting a client-server architecture and leveraging established design patterns such as the Singleton and Factory patterns, the system demonstrates resilience, scalability, and efficiency in managing client connections and facilitating communication. The use of TCP sockets ensures reliable data transmission, while the incorporation of fault tolerance mechanisms enhances the system's ability to withstand client failures and disruptions gracefully.

In order to further increase the capabilities and performance of the networked distributed system, there are a number of directions that future research and development might go. First, investigating ways to reduce overheads and maximize resource usage while managing client connections may result in more effective use of system resources and enhanced scalability. Furthermore, the incorporation of sophisticated communication protocols or technologies, like asynchronous messaging frameworks or WebSockets, may provide improved real-time communication capabilities and support for a wider range of use cases. Advanced security features including encryption and authentication procedures would also guarantee the secrecy and integrity of communication channels while fortifying the system's resistance to external security threats.

Overall, the networked distributed system developed in this study lays a solid foundation for building scalable and reliable communication systems in distributed computing environments. By addressing the challenges of communication coordination and fault tolerance, the system demonstrates its suitability for a wide range of applications, from collaborative platforms to IoT (Internet of Things) networks. Through continued research and development efforts, the system can evolve to meet the ever-growing demands for efficient and secure communication in the digital age, contributing to the advancement of distributed computing technologies and their practical applications.

## Acknowledgement

We would like to extend our gratitude to our tutor, Taimoor Khan, for their invaluable guidance and support throughout the duration of this coursework. Their expertise and insights have been instrumental in shaping the direction of this project and ensuring its successful completion.

Furthermore, we acknowledge the collective effort put forth by our group, which has been essential in overcoming challenges and ensuring the successful execution of the project. Each member's unique skills and perspectives have significantly contributed to the overall quality and effectiveness of our work.

Together, we have demonstrated the power of collaboration and teamwork, underscoring the importance of collective effort in achieving shared objectives. We are grateful for the opportunity to collaborate with such talented individuals and look forward to future endeavors together.